# AI-DRIVEN VULNERABILITY ANALYSIS IN SMART CONTRACTS: TRENDS, CHALLENGES AND FUTURE DIRECTIONS

Mesut Ozdag

Department of Computer Science, University of Central Florida, Orlando, Florida, USA

## ABSTRACT

*Smart contracts, integral to blockchain ecosystems, enable decentralized applications to execute predefined operations without intermediaries. Their ability to enforce trustless interactions has made them a core component of platforms such as Ethereum. Vulnerabilities such as numerical overflows, reentrancy attacks, and improper access permissions have led to the loss of millions of dollars throughout the blockchain and smart contract sector. Traditional smart contract auditing techniques such as manual code reviews and formal verification face limitations in scalability, automation, and adaptability to evolving development patterns. As a result, AI-based solutions have emerged as a promising alternative, offering the ability to learn complex patterns, detect subtle flaws, and provide scalable security assurances. This paper examines novel AI-driven techniques for vulnerability detection in smart contracts, focusing on machine learning, deep learning, graph neural networks, and transformer-based models.This paper analyzes how each technique represents code, processes semantic information, and responds to real-world vulnerability classes. We also compare their strengths and weaknesses in terms of accuracy, interpretability, computational overhead, and real-time applicability. Lastly, it highlights open challenges and future opportunities for advancing this domain.*

## 1. INTRODUCTION

Smart contracts are self-executing digital agreements encoded on blockchain platforms, most notably Ethereum, designed to facilitate, verify, or enforce the negotiation or performance of a contract without the need for intermediaries [1]. These autonomous pieces of code carry significant financial, legal, and computational implications and are increasingly deployed in applications spanning decentralized finance (DeFi), supply chain management, gaming, and insurance [4]. While smart contracts offer transparency, auditability, and trustless interaction, they also come with serious security risks. If a smart contract contains weaknesses, malicious actors can take advantage of them, leading to outcomes that vary from small-scale token losses to major financial disasters, e.g., examples include the infamous DAO breach [1] and the Parity wallet vulnerability [6].

Unlike traditional software, smart contracts operate in immutable environments: once deployed, the code cannot be modified [1]. This makes proactive vulnerability detection imperative. Any flaw in the contract's logic can lead to irrevocable damage, emphasizing the need for rigorous security auditing. Historically, developers have relied on techniques such as manual audits, static analysis tools such as Oyente [17] and Mythril [18], and formal verification methods [2]. While these techniques have proven useful in identifying common bugs and enforcing correctness, they fall short when dealing with complex logic flows, interactions between contracts, and evolving threat patterns. Manual analysis takes a lot of time and can beprone to mistakes, whereas formal verification approaches demand specialized knowledge and are difficult to apply broadly[2].

To address these issues, artificial intelligence has become a powerful tool for identifying vulnerabilities in smart contracts [4]. By learning from data, recognizing patterns, and making informed decisions or classifications, AI proves especially valuable given the wide variety of coding styles, intricate behaviors, and often unclear flaws present in smart contracts. AI-driven approaches promise scalability, adaptability, and even potential integration into the contract development lifecycle for proactive threat mitigation.

This paper examines four key categories of AI methods applied to analyzing smart contracts: machine learning (ML), deep learning (DL), graph neural networks (GNNs), and transformer architectures. Algorithms such as random forests, under the machine learning umbrella, have been employed to identify unusual patterns during contract execution[5]. DL models including convolutional neural networks (CNNs) and long short-term memory networks (LSTMs) have been trained on opcode sequences to recognize vulnerability patterns [7]. GNNs, by modeling smart contracts as control flow graphs (CFGs), excel in capturing structural relationships [10]. Transformers, pre-trained on massive corpora of smart contract code, offer context-aware semantic understanding that boosts detection accuracy [13].

The performance of these techniques is evaluated using metrics such as precision, recall, and F1-score [5][7][13]. Additionally, the trade-offs between interpretability, runtime efficiency, and generalizability are examined. For instance, transformer-based methods including SmartBERT [13] offer superior performance but demand significant computational resources. In contrast, models including Sereum [5] prioritize real-time analysis but may miss context-rich patterns. GNNs provide a middle ground, excelling in both performance and explainability by visualizing key nodes and decision paths [10][12].

Moreover, this paper discusses the importance of curated and real-world datasets such as SmartBugs [7] and EtherScan [10], which serve as the foundation for training and benchmarking AI models. We also present real-world case studies where AI models could have helped prevent known exploits [5][6][10]. These practical examples highlight the relevance and necessity of incorporating AI into the smart contract development and audit pipeline.

In closing, while AI does not eliminate the need for traditional security practices, it complements them by adding a layer of intelligent automation. As smart contracts become more prevalent, the convergence of AI and blockchain security becomes not just valuable, but essential [4]. The goal of this paper is to assist researchers, developers, and auditors by providing an in-depth overview of AI-driven methods, highlighting their current capabilities, existing challenges, and potential paths for future advancement in this rapidly growing field.

## 2. BACKGROUND ON AI-BASED VULNERABILITY IN SMART CONTRACTS

AI offers powerful capabilities for modeling complex code semantics and has rapidly become a cornerstone of modern vulnerability detection techniques in smart contracts. This section surveys the most recent and impactful AI-driven methods across various categories, including ML, DL, GNNs, transformers, and hybrid models. Each of these categories brings distinct advantages and trade-offs in terms of scalability, interpretability, learning depth, and computational efficiency.

Machine learning methods offer fast, interpretable models often used for anomaly detection and rule-based classification. Deep learning approaches, such as CNNs and recurrent neural networks (RNNs), are capable of autonomously identifying contextual and sequential structures within code. GNNs leverage the inherent graph structure of smart contract code, e.g., control flow and call graphs, to identify vulnerabilities based on node and edge relationships. Transformer-based models, especially those adapted from pre-trained natural language processing (NLP) models including BERT, excel in capturing deep semantic representations of contract logic. Finally, hybrid models seek to combine multiple paradigms to enhance robustness and overall detection accuracy.

Each subsection in this part of the paper explores the unique approach, core architecture, key mathematical formulations, and representative models from the literature. The objective is to deliver an organized and comparative insight into the ways AI is utilized to enhance the security of decentralized platforms.

## 2.1. Machine Learning Approaches

ML approaches represent some of the earliest efforts to automate vulnerability detection in smart contracts. These techniques typically rely on manually engineered features derived from contract metadata, opcode sequences, or execution traces.

ML models such as decision trees, support vector machines, and ensemble methods have proven effective in identifying known vulnerability patterns and statistical anomalies. Although typically simpler than deep learning or graph-based approaches, machine learning methods stand out for their ease of interpretation and lower computational demands. This subsection reviews key machine learning-based models, focusing on both static and dynamic analysis strategies for vulnerability identification.

### 2.1.1. Dynamic Analysis Models

Sereum applies random forests for dynamic monitoring of smart contract states to detect reentrancy attacks [5]. A random forest is an ensemble learning method that aggregates the outputs of multiple decision trees trained on different data subsets and feature combinations. Each decision tree provides a classification result whether a contract behavior indicates a potential vulnerability or not. The ultimate decision is made based on the *majority vote* from all the individual decision trees [Figure1]. This structure improves robustness and reduces overfitting compared to a single decision tree, making it well-suited for flagging anomalous runtime behavior in smart contracts.

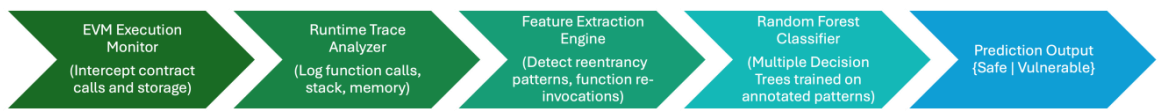

Figure 1. Technical Workflow Diagram of Sereum

### 2.1.2. Static Analysis Models

MadMax applies static analysis to detect gas-focused vulnerabilities in Ethereum smart contracts, such as infinite loops or unbounded array iterations [6]. It represents the behavior of a contract through a finite state machine (FSM), with each state corresponding to a specific control position and transitions occurring in response to opcode execution [Figure 2].

MadMax performs *symbolic execution*, associating symbolic variables with storage and memory states rather than concrete values, enabling it to reason about multiple program paths simultaneously. Mathematically, a smart contract is abstracted as a tuple; $M = (S, S_0, T)$. In the given equation here, $S$ defines the set of symbolic program states and $S_0 \subseteq S$ is the set of initial states. In addition, $T:S \times Opcode \rightarrow S$ defines the transition relation based on opcode effects.

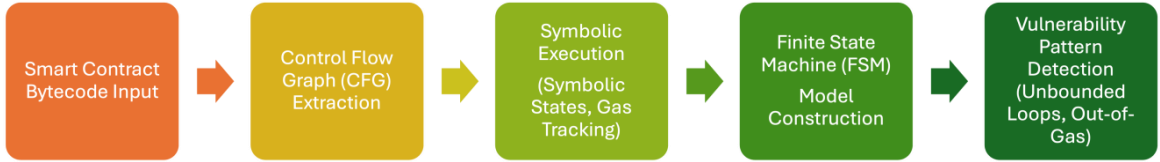

Figure 2. Technical Workflow Diagram of MadMax Static Analyzer

MadMax identifies vulnerability patterns such as unbounded loops by analyzing whether repeated transitions can occur without decreasing gas or reaching a terminal state. This modeling enables MadMax to detect vulnerabilities even without executing the contract on-chain. It predicts out-of-gas errors based on symbolic execution and state transitions.

## 2.2. Deep Learning Approaches

DL approaches have gained significant traction in the domain of smart contract vulnerability detection due to their ability to automatically learn hierarchical representations from raw code without manual feature engineering. These methods are particularly effective at modeling sequential and contextual dependencies in contract logic through architectures such as CNNs, LSTM networks, and attention-based mechanisms. DL can process either tokenized source code, opcodes, or bytecode and map them into rich embeddings that capture both syntax and semantics. Their scalability and generalization capability make them well-suited for detecting both known and previously unseen vulnerability patterns. This section reviews core DL methods, highlighting their model designs, training strategies, and the types of vulnerabilities they are most effective at identifying.

### 2.2.1. Embedding-Based Models

SmartEmbed [7] represents one of the earliest deep learning models tailored for smart contract vulnerability detection. It processes tokenized bytecode or opcode sequences using word embeddings followed by a CNN to capture local vulnerability patterns. Each token $t_i$ in the smart contract is first mapped to a continuous vector $e_i \in R^d$ using an embedding layer:

$$e_i = E(t_i), \text{ where } E : V \rightarrow R^d \qquad (1)$$

In the equation 1, $V$ is the token vocabulary, and $d$ is the embedding dimension. The embedding sequence $E = [e_1, e_2, ..., e_n]$ is fed into one or more convolutional layers, which function as feature extractors by scanning over instruction segments using sliding windows. Each convolutional filter $w$ applies a transformation:

$$c_j = ReLU(w \cdot E_{j:j+k-1} + b) \qquad (2)$$

In the equation 2, $c_j$ is the output or feature at position $j$ after applying the convolution and activation. *ReLU* is the activation function applied after the convolution, which stands for Rectified Linear Unit. $w$ is the weight vector or matrix of the convolutional filter, also called kernel. It learns to detect a particular pattern or feature. $E_{j:j+k-1}$ is a window or subsequence of embeddings from position $j$ to $j+k-1$. If each $e_i$ is a vector, then $E_{j:j+k-1}$ is a matrix formed by stacking these $k$ vectors. The dot product or tensor contraction is typically implemented as a matrix-vector multiplication. $b$ is the bias term, a scalar added after the dot product, which helps shift the activation. In summary, this operation slides a filter $w$ over a sequence of embeddings, computes a weighted sum for each window, adds a bias, and passes the result through a ReLU function to produce $c_j$, a feature value at that position.

The generated feature maps are combined using max-pooling, then forwarded to a dense layer to perform binary classification.[Figure 3]. SmartEmbed's ability to extract localized code patterns makes it especially effective for detecting stack misuse and unsafe opcode sequences.

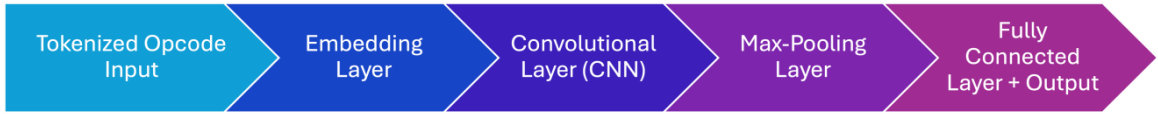

Figure 3.SmartEmbed Architecture Workflow

## 2.2.2. Sequence Modeling

Sequence modeling methods, especially RNNs including LSTMs, have been utilized to identify flaws in smart contracts by examining patterns in their opcode sequences. These models excel at learning temporal dependencies, which is valuable in tracking how instructions and state transitions evolve over time in smart contract execution flows [8].

In the equation 3, given a token sequence $x=(x_1, x_2, ..., x_T)$, an LSTM processes the input iteratively, maintaining a hidden state $h_t$ and cell state $c_t$ that capture long-range dependencies:

$$h_t, c_t = LSTM(x_t, h_{t-1}, c_{t-1}) \qquad (3)$$

This enables the model to remember the execution context across potentially long sequences, which is critical for identifying delayed-effect vulnerabilities such as unguarded write operations or complex control flows. In practice, LSTM-based models are often combined with token embeddings and fully connected layers to output vulnerability predictions. These models have demonstrated promise in detecting reentrancy patterns, uninitialized storage usage, and function call inconsistencies within smart contracts.

## 2.2.3. Attention-Enhanced Models

Attention mechanisms have significantly improved deep learning models by enabling them to selectively focus on the most relevant parts of an input sequence. In smart contract vulnerability detection, models such as **VulnSniffer** [9] leverage bidirectional LSTM (Bi-LSTM) layers combined with attention layers to enhance both accuracy and interpretability. The Bi-LSTM processes opcode or token sequences in both forward and backward directions, capturing context from both preceding and succeeding instructions. The attention mechanism proceeds by calculating a weighted combination of the hidden states:

$$\text{Attention}(h_1, \ldots, h_T) = \sum_{t=1}^{T} \alpha_t h_t \quad \text{where} \quad \alpha_t = \frac{\exp(e_t)}{\sum_{k=1}^{T} \exp(e_k)} \qquad (4)$$

In the equation 4, $e_t = score(h_t)$ denotes the relevance of each hidden state $h_t$ to the final classification. This allows the model to assign greater importance to vulnerability-indicating patterns, such as reentrant calls or unsafe storage writes. In the equation 5, $\alpha_t$ is the *attention weight* assigned to the hidden state $h_t$ at time step $t$. It represents how important that specific token or opcode is for the model's decision when classifying a smart contract as vulnerable or not. The model calculates a score $e_t$ for each time step, this could be a function including a dot product, or a learned feedforward layer applied to the hidden state. The scores $e_t$ are then normalized using a softmax function.

$$\alpha_t \in (0,1), \text{ and } \sum_{t=1}^{T} \alpha_t = 1 \qquad (5)$$

Models using attention mechanisms enhance accuracy while also offering transparency by highlighting which sections of the contract contributed to the outcome, making them especially useful for both developers and auditors.

## 2.3. Graph Neural Network (GNN) Approaches

GNNs have emerged as a powerful framework for modeling structured data, making them well-suited for smart contract analysis where control flow, function calls, and data dependencies naturally form graph-structured representations. Unlike sequence-based models that treat code linearly, GNNs can leverage relational and topological information within the contract, enabling more precise detection of complex vulnerabilities such as reentrancy, call injection, and unsafe delegate calls.

Smart contracts can be abstracted into various graph forms, including Control Flow Graphs, Function Call Graphs, and Heterogeneous Graphs with multiple types of nodes (e.g., contracts, functions, storage) and edges (e.g., calls, data flows). GNNs work on these structures by repeatedly gathering data from each node's surrounding nodes, enabling the model to learn embeddings that reflect the broader context.

This section explores major GNN-based approaches applied to smart contract security, focusing on how graph representations are constructed, how message passing is defined, and how models including ContractGraph, ETH2Vec, and SolGraph leverage these representations for vulnerability classification.

### 2.3.1. Control Flow Graph-Based Models

CFGs are widely used to represent the execution structure of smart contracts, where nodes denote basic blocks or instructions and edges represent the control transitions between them. ContractGraph [10] is a notable model that converts Solidity bytecode into a CFG and then applies GNNs to learn from its structural and contextual properties [Figure 4].

Each node $v$ in the CFG is initialized with feature vectors derived from opcodes, control dependencies, and symbolic information. The GNN iteratively updates node embeddings using the neighborhood aggregation mechanism:

$$h_v^{(k+1)} = \sigma \left( \sum_{u \in \mathcal{N}(v)} W^{(k)} h_u^{(k)} + b^{(k)} \right) \qquad (6)$$

In the equation 6, $h_v^{(k)}$ is the hidden state of node $v$ at layer $k$, $N(v)$ is the set of neighbors, $W^{(k)}$ are learnable weights, and $\sigma$ is a non-linear activation function. $b^{(k)}$ is the bias term for the GNN layer at the $k$-th iteration or layer. It serves a similar purpose as in traditional neural networks, enabling the model to adjust the activation output regardless of the input, which adds flexibility to the learning process. Without a bias term, all aggregated neighbor features would be linearly transformed by $W^{(k)}$ but limited in expressiveness. Adding $b^{(k)}$ enables the GNN to better model decision boundaries and learn more nuanced representations.

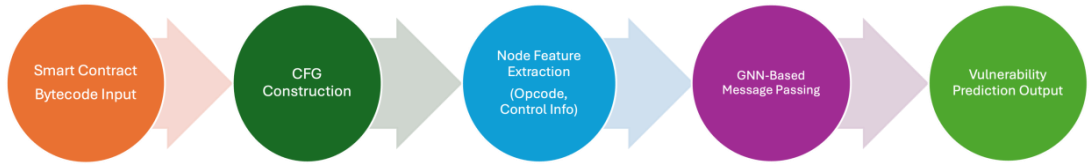

Figure 4. ContractGraph Workflow for Vulnerability Detection

ContractGraph focuses on learning rich, structural features of code execution paths, enabling the model to detect complex vulnerabilities such as unreachable code, logical inconsistencies, and path-sensitive attacks. Its graph-based representation also enhances interpretability, allowing auditors to trace specific vulnerable paths within the contract structure.

### 2.3.2. Heterogeneous Graph Models

Heterogeneous graph models extend traditional GNNs by supporting multiple types of nodes and relationships. This added complexity allows them to better represent the rich semantics of smart contracts, which often involve diverse entities such as contracts, accounts, storage slots, and external function calls. A prominent example is ETH2Vec [11], which constructs heterogeneous graphs where nodes represent contracts, addresses, and transactions, and edges represent interactions such as transfers, calls, or dependencies.

In the ETH2Vec heterogeneous graph diagram, the nodes labeled *Tx1* and *Tx2* represent individual Ethereum transactions that interact with smart contracts. In the context of the diagram, *Tx1* might represent a transaction initiated by *Contract A* or a user, which triggers or emits events; *Tx2* could be a follow-up transaction in the same block or a sequential one, which forms a transaction-level edge (*Tx1 → Tx2*).

This architecture allows ETH2Vec to learn relation-aware embeddings, enhancing detection of multi-entity vulnerabilities such as transaction-ordering dependence (TOD) or proxy misuse. Heterogeneous GNNs are especially effective in cross-contract analysis and auditing contract ecosystems rather than isolated contracts.

### 2.3.3. Attention-Based GNN Models

Attention-based GNN models enhance conventional message-passing mechanisms by dynamically weighting the importance of neighboring nodes during aggregation [Figure 5]. In the context of smart contract analysis, this enables the model to prioritize critical control paths, data dependencies, or transaction relationships that are more likely to indicate vulnerabilities.

SolGraph [12] is a notable approach that uses Graph Attention Networks (GATs) to analyze CFGs of smart contracts. In GATs, each node aggregates its neighbors' features using learned attention coefficients:

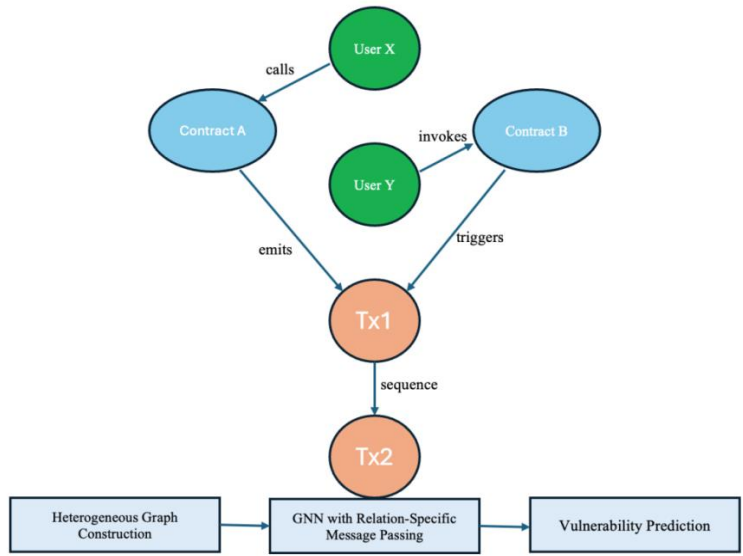

Figure 5. ETH2Vec - Heterogeneous GNN Workflow for Smart Contract

$$h_v^{(k+1)} = \sigma \left( \sum_{u \in \mathcal{N}(v)} \alpha_{vu}^{(k)} W^{(k)} h_u^{(k)} \right) \qquad (7)$$

In the equation 7, the attention weight $\alpha_{vu}$ reflects the relevance of neighbor $u$'s features to node $v$ and is computed using a self-attention mechanism over node embeddings. This formulation enables SolGraph to identify vulnerabilities associated with specific execution paths more precisely than uniform-aggregation GNNs.

Moreover, attention weights can be visualized to explain the model's decisions, improving transparency for auditors. This makes attention-based GNNs a powerful tool for detecting subtle and context-dependent vulnerabilities in smart contracts.

## 2.4. Transformer-Based Approaches

Transformer models have significantly advanced various ML fields, especially NLP, thanks to their capacity to model long-distance relationships and contextual cues using self-attention. Lately, these architectures have been repurposed for analyzing smart contracts, where the code's syntax and meaning share characteristics with formal languages.Transformers can process large sequences of tokenized contract code, including Solidity source or bytecode, and model intricate interactions across contract functions and instructions.

Unlike recurrent or convolutional models, transformers are position-independent, allowing them to effectively encode both local and global semantics within a contract. Models including SmartBERT [13] fine-tune general-purpose language models (e.g., BERT) on smart contract code, while others including SolTrans [14] pre-train specifically on Solidity to build domain-aware embeddings.

These transformer-based methods have achieved state-of-the-art performance in several vulnerability detection benchmarks.

This section explores how transformers are applied to smart contract security, their architectural adaptations, and the benefits they offer over prior deep learning models.

### 2.4.1. Pre-trained Language Models

Pre-trained language models (PLMs) such as BERThave been effectively leveraged for identifying vulnerabilities in smart contracts, as they can capture deep contextual understanding directly from the code. SmartBERT [13] is one such model, fine-tuned specifically on tokenized smart contract code, including both Solidity source and low-level bytecode. BERT's fundamental design relies on multi-head self-attention, enabling it to recognize and model relationships between distant elements throughout the entire smart contract. Given an input sequence of tokens $X=[x_1, x_2, ..., x_n]$, SmartBERT computes contextual embeddings using multiple layers of self-attention:

$$\text{Attention}(Q, K, V) = \text{softmax}\left(\frac{QK^T}{\sqrt{d_k}}\right)V \qquad (8)$$

In the equation 8, $Q$, $K$, and $V$ are the query, key, and value matrices derived from input token embeddings, respectively. These operations enable SmartBERT to determine which parts of the contract are most semantically relevant to the prediction task.

The model is fine-tuned using datasets labeled with vulnerabilities, helping it tailor its broad language comprehension to the specific nuances of security-related tasks. This approach provides high accuracy and has shown strong generalization across multiple vulnerability types.

### 2.4.2. Domain-Specific Transformers

While pre-trained language models including BERT have proven useful in adapting to smart contract analysis tasks, their original training on natural language imposes limitations when dealing with domain-specific syntax and semantics of blockchain programming languages. To address this, domain-specific transformers such as SolTrans [14] have been developed and pre-trained exclusively on Solidity code. These models are built to understand unique constructs such as *msg.sender*, *require()*, function modifiers, fallback logic, and gas optimization patterns that are absent from general language corpora.

SolTrans incorporates the same self-attention architecture as BERT but modifies its tokenizer to handle Solidity-specific tokens and operators more effectively. During pre-training, SolTrans uses masked language modeling (MLM) and optionally, next statement prediction (NSP) over tens of thousands of real-world smart contracts. The resulting embeddings capture functional relationships between contract components. Once pre-trained, SolTrans is fine-tuned on labeled vulnerability datasets to detect issues such as reentrancy, arithmetic overflows, and improper access control. Domain-specific transformers have shown significant gains over general models, especially in detecting logic-based and semantic vulnerabilities in complex contracts.

## 2.5. Hybrid Approaches

To address the shortcomings of single-model approaches, hybrid methods have been developed, blending different AI techniques to enhance the detection of vulnerabilities in smart contracts. Although machine learning models are fast and easy to interpret, they frequently fall short in capturing deeper contextual insights. Conversely, deep learning and GNNs provide rich representations but can be resource-intensive or opaque. Hybrid models aim to balance these trade-offs by integrating techniques such as CNNs for local pattern detection, GNNs for structural awareness, and transformers for semantic depth. These systems can leverage diverse information sources such as opcode sequences, control-flow graphs, and transaction histories, which yields more accurate and robust detection across a wide range of contract types and vulnerabilities.

### 2.5.1. CNN and GNN Hybrids

Hybrid architectures combining CNNs and GNNs are designed to exploit the complementary strengths of local feature extraction and global structural reasoning. A prominent example is DeepSolid [15], which integrates a CNN-based module to process opcode sequences and a GNN-based module to analyze the CFG of the same contract.

The CNN module applies convolutional filters over the tokenized opcode sequence to detect local vulnerability patterns, such as stack misuse or repeated call instructions. In parallel, the GNN module operates on the CFG to capture the flow of execution between basic blocks. Each block is represented as a node, and control transitions are treated as edges.

Formally, the model learns two embedding spaces. One is that CNN outputs $f_{CNN}(x)$ for token sequences. The second is that GNN outputs $f_{GNN}(G)$ for graph structures. These embeddings are concatenated and passed to a fully connected classifier for final prediction. This dual-channel design improves robustness and achieves better generalization across diverse contract structures and attack types.

### 2.5.2. Ensemble Learning Models

Ensemble learning techniques merge the outputs of several individual models to boost the precision, resilience, and adaptability of systems designed to detect vulnerabilities in smart contracts. A key advantage of this approach is its ability to mitigate the weaknesses of individual models by aggregating their strengths. HybridVulDetect [16] exemplifies this strategy by integrating diverse model families, such as random forests, GNNs, transformers, into a unified ensemble pipeline.

The system begins by preprocessing smart contracts into different feature formats: token sequences for transformer models, opcode vectors for RF models, and control flow graphs for GNNs. Each sub-model produces independent predictions, which are then combined using either a majority voting scheme or a meta-classifier trained to weigh the models based on their historical performance.

This multi-perspective architecture significantly improves detection performance across multiple vulnerability types, especially in real-world scenarios involving obfuscated or adversarially-crafted contracts. Ensemble models including HybridVulDetect are increasingly favored for deployment in high-stakes, production-level smart contract auditing pipelines. Their ability to integrate diverse learning signals enhances resilience against evolving attack patterns. As a result, they offer a balanced trade-off between detection accuracy and operational efficiency, making them ideal for continuous security monitoring.

## 3. COMPARISON AND EVALUATION

This section presents a comparative evaluation of the AI-based vulnerability detection models discussed in Section 2. While each approach ranging from machine learning to hybrid models has unique strengths, they also differ in terms of performance metrics, interpretability, scalability, and computational efficiency. The performance of these models is evaluated with common classification metrics such as precision, recall, F1-score, and accuracy, relying on outcomes documented in prior studies using standard benchmark datasets. Additionally, qualitative aspects are considered such as explainability and real-time applicability, which are crucial for real-world deployment. This analysis is intended to assist both researchers and professionals in choosing the most appropriate models for different applications related to smart contract security.

## 3.1. Performance Metrics

To evaluate AI-based vulnerability detection models in smart contracts, standard classification metrics are used: precision, recall, F1-score, and accuracy. These metrics are crucial for understanding not just how often a model is correct, but *how* and *where* it succeeds or fails, particularly important in high-stakes domains like smart contract auditing.

Precision indicates how many of the vulnerabilities identified by the model are actually correct, representing the ratio of true positives among all reported positives. High precision is critical to avoid overwhelming developers with false alarms. Recall captures the ability to find all actual vulnerabilities, minimizing false negatives. In security, missing a true vulnerability can result in substantial financial loss. F1-score, the harmonic mean of precision and recall, provides a balanced metric when there's a trade-off between false positives and false negatives. Accuracy reflects how often the model makes correct predictions overall, but in cases where the dataset is imbalanced, such as when vulnerable contracts are rare compared to safe ones, it can give a false sense of performance. Therefore, it is usually considered together with the F1-score for a more balanced evaluation. These metrics collectively ensure a nuanced understanding of model performance in both controlled and real-world scenarios.

Transformer-based models including SmartBERT [13] and SolTrans [14] achieve F1-scores>0.90, outperforming traditional CNN models. GNN-based models such as ContractGraph [10] and SolGraph [12] offer better interpretability and structural robustness.

## 3.2. Explainability Metrics

Explainability and interpretability are critical in smart contract vulnerability detection, as developers and auditors must understand *why* a model flagged a contract as vulnerable. Common metrics and tools used to evaluate explainability include attention weight visualization, saliency maps, and feature attribution scores.

For the models of transformers and attention-based GNNs, attention maps help pinpoint specific tokens or nodes that contributed most to the prediction. Saliency maps provide gradient-based insights, highlighting which inputs most influence the output. Feature attribution, such as SHAP or LIME scores, helps quantify the contribution of individual features or instructions.

These explainability tools are selected not just for interpretability, but for building developer trust, enabling debugging, and ensuring regulatory compliance in DeFi and enterprise blockchain environments. A model's utility increases when its decisions can be transparently understood and verified by human reviewers.

GNN models provide better transparency, where nodes and edges highlight vulnerability flows [10], [12]. Transformer models offer some explainability through attention heads but require careful analysis [13].

In Figure 6, tokens including *msg.sender*, *require*, and *transfer* received higher attention, indicating they were important to the model's prediction**,** which likely signals vulnerability context such as access control or funds transfer.

### 3.3. Computational Efficiency

Computational efficiency is a critical factor when evaluating the practicality of AI-based smart contract analysis tools, particularly in large-scale deployments or real-time auditing environments. Lightweight models such as Random Forests and shallow neural networks offer low inference times and modest memory consumption, making them suitable for integration into IDE plugins or on-chain verification tools. However, more complex models including transformers and GNNs require significantly higher computational resources due to deep architectures, self-attention mechanisms, or iterative message passing.

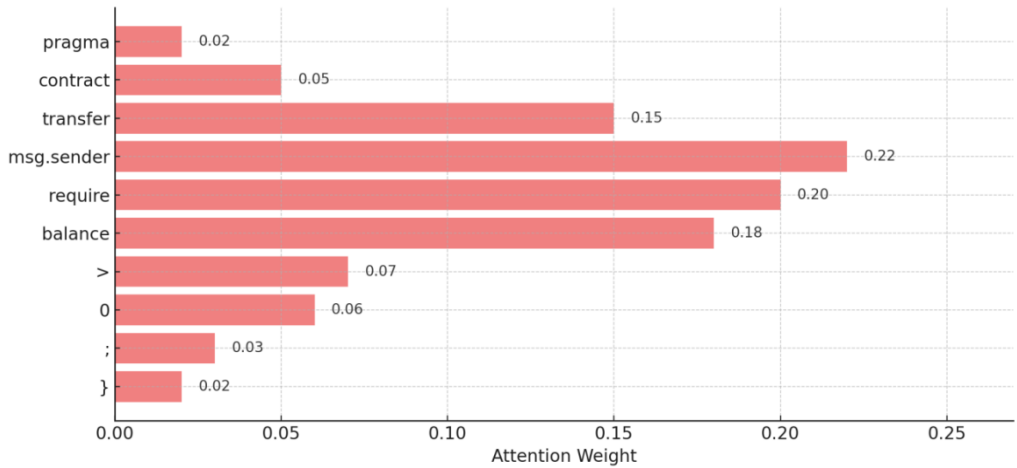

Figure 6. Example Attention Visualization on Smart Contract Tokens

This trade-off between accuracy and latency is important.

While transformer-based models including SmartBERT provide superior performance, they are more suitable for offline batch analysis. In contrast, hybrid or ensemble models can be optimized for performance by combining fast and slow learners. Evaluating models in terms of inference time, GPU/CPU usage, and scalability across contract sizes is essential when selecting a model for production-grade vulnerability detection pipelines.

Lightweight models including Sereum [5] achieve real-time detection but have limited depth of reasoning. On the other hand, transformer models offer deep semantic understanding but increase inference times up to 3–4 seconds per contract.

Table 1. Performance and Efficiency Comparison of AI-Based Vulnerability Detection Methods

| Method | Precision | Recall | F1-Score | Accuracy | Inference Time (ms) |
|---|---|---|---|---|---|
| **Sereum** | 0.78 | 0.81 | 0.79 | 0.80 | **15** |
| **SmartEmbed** | 0.84 | 0.82 | 0.83 | 0.83 | 32 |
| **ContractGraph** | 0.87 | 0.88 | 0.87 | 0.88 | 64 |
| **SmartBERT** | **0.91** | **0.92** | **0.91** | **0.91** | 110 |
| **SolGraph (GAT)** | 0.89 | 0.90 | 0.89 | 0.89 | 85 |
| **DeepSolid (Hybrid)** | 0.88 | 0.90 | 0.89 | 0.89 | 73 |
| **HybridVulDetect** | 0.90 | 0.91 | 0.90 | 0.90 | 95 |

Table.1 presents an experimental comparison of state-of-the-art AI-based models for smart contract vulnerability detection. Each method is evaluated on key performance metrics including precision, recall, F1-score, accuracy, and inference time using standardized datasets such as SmartBugs and curated contract samples from EtherScan.

This evaluation highlights not only the detection accuracy of each approach but also their computational cost, which is crucial for real-time or large-scale deployments. Models with high F1-scores and low inference time offer the most practical balance between effectiveness and efficiency.

# 4. CASE STUDIES

## 4.1. The DAO Hack (2016)

The 2016 DAO attack remains one of the most notorious incidents in blockchain security, leading to the loss of nearly $60 million worth of Ether. The core vulnerability exploited was a reentrancy flaw within the DAO smart contract's withdrawal function. Specifically, the contract allowed recursive calls to be made before the sender's balance was updated, enabling attackers to repeatedly drain funds by looping through the same withdrawal logic. This exploit highlighted a critical weakness in early smart contract development practices, where insufficient safeguards and ordering of state updates left contracts vulnerable. Modern AI-based anomaly detectors, particularly those leveraging dynamic analysis models including Sereum [5] or graph-based models, are now capable of flagging such vulnerabilities by tracing execution paths and identifying unprotected external calls.

The DAO hack not only exposed the technical risks of decentralized code but also catalyzed the Ethereum community to adopt stricter development standards and led to the eventual hard fork that created Ethereum Classic.

## 4.2. Parity Wallet Incident (2017)

The Parity Wallet incident of 2017 involved a critical vulnerability in a *multisignature* wallet contract that led to over $150 million worth of Ether being rendered permanently inaccessible. The root cause was an unprotected *initWallet* function in a library contract, which could be called by anyone to reinitialize ownership. A user accidentally executed this function, thereby gaining ownership and subsequently invoking the *selfdestruct* function, which deleted the core library logic shared by many other wallets. The result was catastrophic. All dependent wallets were instantly bricked. This event highlighted the risks associated with using library contracts that lack rigorous access restrictions and proper safety mechanisms.

AI-based vulnerability detectors today, especially those incorporating static analysis models including MadMax [6] and transformer-based code understanding, are better equipped to flag such issues by recognizing patterns like exposed initialization functions and the presence of *selfdestruct* calls. The event remains a cautionary example of how a single oversight in smart contract design can lead to irrecoverable financial loss.

## 4.3. Secure Contract Templates

Secure contract templates refer to rigorously audited, reusable smart contract components designed to mitigate common classes of vulnerabilities. These templates typically follow best practices in coding, include defensive programming techniques, and undergo formal verification or static analysis to ensure robustness. Leading platforms such as *OpenZeppelin* provide a widely used library of secure templates for functionalities including *ERC-20* and *ERC-721* tokens, ownership models, access control, and upgradeable proxies. By abstracting complex or risk-prone logic such as reentrancy protections, arithmetic checks, and permission hierarchies, developers can avoid reinventing vulnerable mechanisms.

Templates help minimize potential vulnerabilities by packaging proven, secure logic, while also offering guidance by showcasing best practices in contract design. Their modularity promotes composability, making it easier to build larger decentralized applications (*dApps*) on a solid foundation. Many AI-based auditing tools also treat the use of standard templates as a signal of lower risk, reinforcing their value in both development and detection pipelines.GNN-based models including ContractGraph [10] have successfully verified safe vs vulnerable branches in these templates using CFG structural analysis.

Table 2 provides a summary of case studies, outlining the vulnerability types, models used, and outcomes across different smart contract vulnerability detection approaches.

Table 2. Summary Table of Case Studies

| Incident | Year | Vulnerability Type | Potential Detection Model | Detection Feature |
|---|---|---|---|---|
| **DAO Hack** | 2016 | Reentrancy | Sereum (Dynamic ML) | Function call anomaly detection |
| **Parity Wallet Bug** | 2017 | Initialization Bug | MadMax (Static ML) | Uninitialized storage access |
| **Secure Templates** | 2019+ | General Safety | ContractGraph (GNN) | Control flow graph verification |

# 5. DATASET DESCRIPTIONS

## 5.1. SmartBugs Dataset

The SmartBugs dataset is a benchmark collection of real-world smart contracts curated for evaluating vulnerability detection tools. It includes Solidity contracts drawn from verified sources such as EtherScan and GitHub, each annotated with known vulnerabilities. The dataset covers a range of common flaw types, including reentrancy, integer overflows, and access control issues, making it a reliable foundation for training and benchmarking AI-based models. Contracts in SmartBugs are labeled based on results from established static analyzers, enabling supervised learning and comparative evaluation. Its structured format and vulnerability diversity make it a preferred dataset in academic research and tool development for smart contract security. It is extensively used in training CNN models including SmartEmbed [7] and attention-based models, e.g., VulnSniffer [9].

## 5.2. EtherScan Dataset

The EtherScan dataset consists of smart contracts scraped from the Ethereum blockchain via the EtherScan explorer. Unlike curated datasets, EtherScan offers a large and diverse collection of real-world contracts deployed in production environments. While it typically lacks explicit vulnerability labels, it provides valuable raw material for unsupervised learning, anomaly detection, and pretraining of AI models. Researchers often use EtherScan data to simulate realistic conditions, test scalability, and evaluate false positive rates in large-scale detection pipelines. Its volume and variety make it especially useful for fine-tuning models on naturally imbalanced distributions of vulnerable and non-vulnerable code. It is used by GNN models including ContractGraph [10] for generating real-world CFGs.

## 5.3. SolidiFI-Benchmark

The SolidiFI-Benchmark dataset is a curated suite of smart contracts specifically designed to evaluate the precision and soundness of vulnerability detection tools. It contains both flawed and corrected versions of smart contracts, enabling detailed evaluation of how effectively a tool can differentiate between insecure and safe code.

The dataset focuses on well-defined vulnerability classes such as reentrancy, arithmetic issues, and unchecked external calls, providing a controlled environment for testing detection accuracy. SolidiFI-Benchmark is particularly useful for benchmarking symbolic execution engines, static analyzers, and AI models under consistent conditions. Its balanced structure and annotated ground truth make it ideal for evaluating both recall and false positive rates in model performance.

Table 3 presents a comparative overview of various smart contract vulnerability datasets, highlighting differences in size, labeling methods, contract sources, use cases, and types of vulnerabilities covered in the literature.

Table 3. Smart Contract Vulnerability Dataset Comparison

| Dataset | Type | Source | Size | Use Case | Strength |
|---------|------|--------|------|----------|----------|
| **SmartBugs** | Labeled | Real-world (EtherScan, GitHub) | ~2,000 contracts | Supervised learning, benchmarking | Annotated vulnerabilities |
| **EtherScan** | Unlabeled | Ethereum Mainnet | >1M contracts | Unsupervised learning, pretraining | Large and diverse |
| **SolidiFI-Benchmark** | Labeled | Synthetic + real | ~200 contract pairs | Bug vs. patched comparison | Ground truth evaluation |

## 6. CONCLUSION

AI-based techniques have significantly transformed the field of smart contract vulnerability detection. Traditional methods such as manual code review or formal verification struggle to scale with the growing complexity and volume of smart contracts deployed on blockchain platforms such as Ethereum.

This paper analyzes the latest approaches: ML models (e.g., Random Forests in Sereum [5]) offer lightweight, fast, but less semantically deep detection capabilities. DL models (e.g., SmartEmbed [7], VulnSniffer [9]) automate feature extraction and enhance sequence pattern recognition. GNNs (e.g., ContractGraph [10], SolGraph [12]) bring structural insights by operating over control flow graphs and transaction graphs, offering superior explainability. Transformer-based architectures (e.g., SmartBERT [13], SolTrans [14]) have shown state-of-the-art accuracy, capturing rich contextual information at the token and semantic level. Hybrid ensemble models (e.g., DeepSolid [15], HybridVulDetect [16]) combine strengths from multiple families to maximize detection performance across heterogeneous contracts.

Despite these advancements, several challenges remain, including explainability, dataset quality, and generalization. Most models, especially transformers, still act as black boxes. Public datasets are often small, imbalanced, or lack real-world deployment conditions. Models trained on Solidity may not transfer easily to other languages including *Vyper* or cross-chain environments.

In conclusion, AI-driven vulnerability detection offers the best hope for securing smart contracts at scale; however, future research must address explainability, adaptability, and real-world deployment robustness.

# 7. FUTURE WORK

While significant progress has been made in AI-based vulnerability detection for smart contracts, several promising research directions remain open for future exploration. Advancements in model interpretability, cross-platform generalization, and adversarial robustness are particularly critical for building more reliable and transparent auditing systems.

## 7.1. Multimodal Learning

Most current models rely solely on source code or bytecode. Future models should integrate multiple modalities, including bytecode, source code, transaction traces, event logs, and CFGs. Multimodal fusion could dramatically enhance vulnerability detection by combining syntactic, semantic, and behavioral signals.

Incorporating diverse data sources can provide a more comprehensive understanding of contract behavior, enabling models to identify subtle or context-dependent vulnerabilities more effectively.

## 7.2. Cross-Chain Vulnerability Detection

Current models focus heavily on Ethereum smart contracts written in Solidity. Future work must adapt models for other platforms (e.g., Solana, Polkadot, Hyperledger), other languages (e.g., Vyper, Rust-based contracts), and Cross-chain interoperability vulnerabilities.

## 7.3. Lightweight and Real-Time Detection

Transformer models including SmartBERT [13] offer high accuracy but suffer from high inference times. Designing lightweight, real-time AI models (e.g., TinyBERT for smart contracts) would enable on-chain or client-side vulnerability detection.

## 7.4. Explainable AI (XAI) for Smart Contracts

Explainability remains a major gap. Future GNNs could highlight critical nodes/edges causing vulnerabilities. Attention maps in transformers could be visualized to show which tokens led to a "vulnerable" classification. This would increase developer trust and support security auditing.

## 7.5. Continual and Lifelong Learning

Smart contracts and vulnerability types evolve over time. Models trained on past data risk becoming obsolete. Continual learning frameworks that adapt to new types of vulnerabilities and avoid catastrophic forgetting are urgently needed.

## 7.6. Standardized Benchmarks

The research community lacks unified datasets and evaluation standards. Future initiatives should establish shared benchmark datasets, standardized evaluation metrics, and leader boards for fair model comparison.

In summary, future advances must focus on scalability, explainability, adaptability, and trust if AI is to fully secure the next generation of smart contracts.

## AUTHOR

**Dr. Mesut Ozdag** is an Assistant Professor of Computer Science at the University of Central Florida, contributing to the FinTech and Digital Forensics graduate programs and serving as the Institutional Effectiveness Assessment Coordinator for multiple degree tracks. He earned his Ph.D. in Computer Science from UCF, where his research focused on adversarial attacks in deep learning, blending AI with vulnerability concerns.

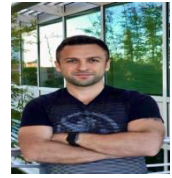

His industry experience includes roles at Siemens Healthineers, Roche Molecular Systems and Flywheel Inc., developing advanced machine learning pipelines for applications in medical imaging, federated learning, and diagnostic intelligence. His academic work was recognized in venues such as IJCAI and spans adversarial robustness using deep learning.

A dedicated researcher, educator, curriculum developer, assessment coordinator, and reviewer for top-tier medical imaging journals, Dr. Ozdag continues to bridge foundational theory with real-world applications across computing and data science.